\documentclass[twocolumn,showpacs,amsmath,amssymb]{revtex4}
\usepackage{graphicx}
\usepackage{dcolumn}
\usepackage{bm}
\newcommand{\be}{\begin{eqnarray}}
\newcommand{\en}{\end{eqnarray}}
\newcommand{\ben}{\begin{eqnarray*}}
\newcommand{\enn}{\end{eqnarray*}}

\newcommand{\bi}{\begin{itemize}}
\newcommand{\ei}{\end{itemize}}

\begin{document}
%%%%%%%%%%%%%%%%%%%%%%%%%%%%%%%%%%%%%%%%%%%%%%%%%%%%%%%%%%%%%%%%%%%%%%%%%%%%%%%%%%%%%%%%%%%%%%%%%%%%%%%%%%%%%%%%%%%%%%%%%%%%
\title{Three-dimensional turbulent relative dispersion by GOY shell model}
%%%%%%%%%%%%%%%%%%%%%%%%%%%%%%%%%%%%%%%%%%%%%%%%%%%%%%%%%%%%%%%%%%%%%%%%%%%%%%%%%%%%%%%%%%%%%%%%%%%%%%%%%%%%%%%%%%%%%%%%%%%%
\author{Sagar Chakraborty}
\email{sagar@nbi.dk}
\affiliation{Niels Bohr International Academy, Blegdamsvej 17, 2100 Copenhagen $\O$, Denmark}
\author{Mogens H.Jensen}
\email{mhjensen@nbi.dk}
\affiliation{Niels Bohr Institute, Blegdamsvej 17, DK-2100 Copenhagen, Denmark}
\author{Bo S. Madsen}
\email{bomadsen@fys.ku.dk}
\affiliation{Niels Bohr Institute, Blegdamsvej 17, DK-2100 Copenhagen, Denmark}
%%%%%%%%%%%%%%%%%%%%%%%%%%%%%%%%%%%%%%%%%%%%%%%%%%%%%%%%%%%%%%%%%%%%%%%%%%%%%%%%%%%%%%%%%%%%%%%%%%%%%%%%%%%%%%%%%%%%%%%%%%%
\date{\today}
%%%%%%%%%%%%%%%%%%%%%%%%%%%%%%%%%%%%%%%%%%%%%%%%%%%%%%%%%%%%%%%%%%%%%%%%%%%%%%%%%%%%%%%%%%%%%%%%%%%%%%%%%%%%%%%%%%%%%%%%%%%%
\begin{abstract}
We study pair dispersion in a three-dimensional incompressible high Reynolds number turbulent flow generated by Fourier transforming the dynamics of the GOY shell model into real space. We show that GOY shell model can successfully reproduce both the Batchelor and the Richardson-Obukhov regimes of turbulent relative dispersion. We also study how the cross-over time scales with the initial separations of a particle pair and compare it to the prediction by Batchelor.
\end{abstract}
%%%%%%%%%%%%%%%%%%%%%%%%%%%%%%%%%%%%%%%%%%%%%%%%%%%%%%%%%%%%%%%%%%%%%%%%%%%%%%%%%%%%%%%%%%%%%%%%%%%%%%%%%%%%%%%%%%%%%%%%%%%%
%%%%%%%%%%%%%%%%%%%%%%%%%%%%%%%%%%%%%%%%%%%%%%%%%%%%%%%%%%%%%%%%%%%%%%%%%%%%%%%%%%%%%%%%%%%%%%%%%%%%%%%%%%%%%%%%%%%%%%%%%%%%
\maketitle
%%%%%%%%%%%%%%%%%%%%%%%%%%%%%%%%%%%%%%%%%%%%%%%%%%%%%%%%%%%%%%%%%%%%%%%%%%%%%%%%%%%%%%%%%%%%%%%%%%%%%%%%%%%%%%%%%%%%%%%%%%%%
\section{Introduction}
Passive particles trajectories in a fluid flow provide an easier way of understanding the nature of the flow than a full-fledged study of the Eulerian field variables of the flow. Recent development in Lagrangian measurement techniques has made it possible to study relative motion of pairs of fluid elements in a turbulent flow \cite{Eberhard,Luthi07}. The understanding of the pair dispersion in turbulence is of importance for comprehending many natural processes such as transport and mixing in natural and engineering flows, formation of warm clouds {\it etc}.
\\
Suppose, $L$ is the integral scale and $\eta$ is Kolmogorov length scale in a turbulent flow, then one can classify the process of dispersion in the flow into three distinct regimes based on the separation of the particles relative to the turbulent scales: $(i)$ dissipation range corresponding to $r(t)\ll\eta$ ($r(t)$ is the separation of a pair of fluid elements or particles at time $t$), $(ii)$ inertial range corresponding to $\eta\ll r(t)\ll L$, and $(iii)$ diffusion range corresponding to $r(t)\gg L$.
From the work of Richardson \cite{Richardson} (and subsequently Obukhov \cite{Obukhov}), in the inertial range of fully-developed three-dimensional turbulence, there exits the well-known Richardson-Obukhov law (R-O law): $\left\langle r(t)^2\right\rangle = g \epsilon t^3$. Here, $\epsilon$ is the energy dissipation rate per unit mass, and $g$ is termed as the Richardson constant. Using similarity arguments, Batchelor \cite{Batchelor} extended the study to both short-time and intermediate-time, and obtained
\begin{equation}
    \left\langle|\textbf{r}(t)^2 - \textbf{r}_{0}^2|\right\rangle = \begin{cases} \frac{11}{3} C_2 \epsilon^{2/3} r^{2/3}_0 t^2 & t\ll t_0\\ g \epsilon t^3 & t\gg t_0 \end{cases},\label{eq1}
\end{equation}
where $C_2$ is the Kolmogorov constant for the longitudinal second-order velocity structure function and
\begin{equation}
t_0 \equiv \left(\frac{r^{2}_{0}}{\epsilon}\right)^{1/3},\label{eq2}
\end{equation}
the classical correlation time of an eddy of size $r_0$, is the time for which a particle-pair ``remembers" their initial separation. $t_0$ is also called Batchelor time. In this paper, we investigated numerically this ``cross-over time" between the two scalings, $\sim t^2$ and $\sim t^3$, and study how it is related to the initial separation $r_0$. In what follows, we denote the first relation in equation ({\ref{eq1}}) as Batchelor's scaling law; the second one is obviously R-O law. Even though we shall be focused on three dimensional fully-developed turbulence, it is worth keeping in mind that R-O is omnipresent \cite{salazar}: there are reports of observing it even in two-dimensional, quasi-two-dimensional and rotating turbulent flows \cite{nicolleau}. Batchelor's law is no less robust--- it has been detected even in turbulent flows that are not perfectly homogeneous and isotropic \cite{Ouellette}.
\\
In the past decade, theoretical researchers have understood --- in significant details --- the process of advection of particles by developed turbulent flows in the framework of Kraichnan model\cite{r1,r2,r3,r4,r5,r6}.
Therein, calculations of the anomalous intermittent exponents for structure functions in the inertial range show that pair structure function, which determines the Richardson scaling exponent, is not intermittent.
\\
Until now, laboratory experiments have neither confirmed nor refuted the existence of R-O scaling. This, of course, is due to the fact that these experiments are unable to attain very high Reynolds number due to practical constraints.
Though environmental experiments can achieve the highest Reynolds numbers, the problem with such experiments are many--- inhomogeneity, anisotropy, rather uncontrollable experimental conditions, vastness of the domain to be studied, {\it etc.}.
Direct numerical simulation of Navier-Stokes equation (DNS), are much more successful in this respect since the computational power is on a rise, but cannot yet compete with the value of Reynolds numbers obtained in shell model simulations. Also, there are issues with the commonly adopted periodic boundary conditions in DNS as they may interfere with the homogeneity of the system.
We believe that shell models are well suited to test the R-O law as not only can one achieve very high Reynolds number  but also there's no sweeping effect disturbing them. It is worth mentioning that sweeping effect's presence makes it hard to observe Kolmogorov scaling \cite{uriel} with which R-O law is so intimately related being the only choice for pair dispersion consistent with Kolmogorov scaling theory. A technique based on a real space transformations of shell models have recently been applied to study force fields study turbulent relative dispersion \cite{luiza} and we adopt a similar method here.
\section{The model}
To study how a pair of particles diffuse in a 3D turbulent flow, we
consider the kinematics of pair particles advected by the
homogeneous turbulent flow obtained by a real-space transformation
of the GOY shell model \cite{luiza,Jensen99,Bohr98}. This model proposed
originally by Gledzer, Yamada and Ohkitani~\cite{Gledzer73,Yamada87}
provides a description of the turbulent motion
embodied in the Navier-Stokes equations. The GOY model is formulated
on a $N$-discrete set of wavenumbers, $k_n=2^n$, with the associated
Fourier modes $u_n$ evolving according to
\begin{eqnarray}
\label{un}
\left(\frac{d}{ dt}+\nu k_n^2 \right) \ u_n \ & = &
 i \,k_n \left(a_n \,   u^*_{n+1} u^*_{n+2} \, + \, \frac{b_n}{2}
u^*_{n-1} u^*_{n+1} \, + \, \right.\nonumber \\
& & \left.\frac{c_n }{4} \,   u^*_{n-1} u^*_{n-2}\right)  \ + \ f \delta_{n,2},
\end{eqnarray}
for $n=1,\cdots, N$. The coefficients of the non-linear terms are
constrained by two conservation laws, namely the total energy, $E =
\sum_n |u_n|^2$, and the helicity (for 3D), $H =
\sum_n(-1)^{n}k_n|u_n|$, or the enstrophy (for 2D), $Z = \sum_n
k_n^2|u_n|^2$, in the inviscid limit, i.e.
$f=\nu=0$~\cite{Biferale95}. Therefore, they may be expressed in
terms of a free parameter only $\delta\in [0,2]$, $ a_n=1,\
b_{n+1}=-\delta,\ c_{n+2}=-(1-\delta)$. As observed by
Kadanoff~\cite{Kadanoff95}, one obtains the canonical value
$\delta = 1/2$, when the helicity is conserved. The set
(\ref{un}) of $N$-coupled ordinary differential equations can be
numerically integrated by standard techniques~\cite{Pisarenko93}. We
have used standard parameters in this paper $N = 27, \nu = 10^{-9},
k_0 = 0.05, f = 5 \times 10^{-3}(1+i)$.

The GOY model is defined in $k$-space but we study particle
dispersion in direct space obtained by an inverse Fourier
transform~\cite{Jensen99} of the form
\begin{equation}
\label{field}
 {\vec v} ({\vec r},t)=\sum_{n=1}^{N}
 {\vec c}_n [u_n(t) e^{i {\vec k}_n\cdot {\vec r}}
+ c.\, c.].
\end{equation}
Here the wavevectors are ${\vec k}_n ~=~ k_n {\vec e}_n$ where
${\vec e}_n$ is a unit vector in a random direction, for each shell
$n$ and ${\vec c}_n$ are unit vectors in random directions. We
ensure that the velocity field is incompressible, $\nabla\cdot{\vec
v} =0$, by constraining ${\vec c}_n \cdot {\vec e}_n =0,~\forall n
$. In our numerical computations we consider the vectors ${\vec
c}_n$ and ${\vec e}_n$ quenched in time but averaged over many
different realizations of these.

\begin{figure}[hbt]
\includegraphics[width=0.50\textwidth]{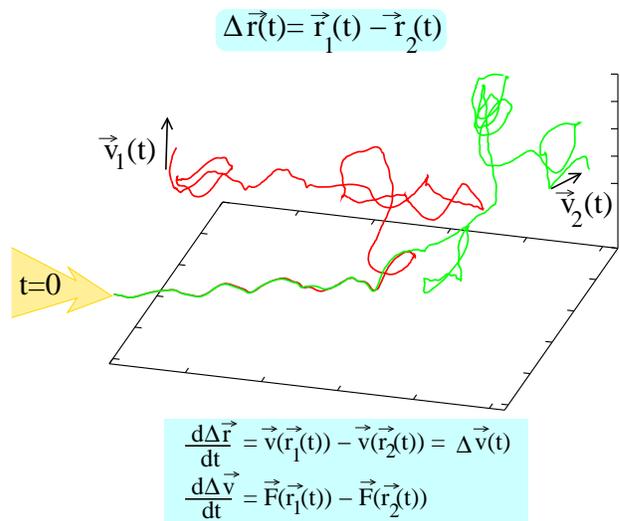}
\caption{({\it Colour Online}) Two particles being advected in a random force
field, generated by the GOY shell model.
\label{goy}}
\end{figure}

As an example of the motion in this field, Fig.~\ref{goy} shows the
trajectories of two passively advected particles. As the relative
distance diverges in time, the two particles experience different
force fields, which in turn typically increase the difference in the
relative velocities of the two particles.  The figure shows the
individual particles as they are advected, first together and later
diverging away from each other when they are encased in different
eddies.
\section{Results}
\begin{figure}[hbt]
\includegraphics[width=0.50\textwidth]{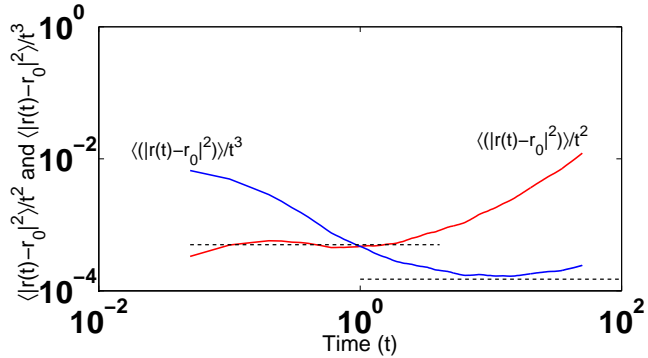}
\caption{({\it Colour online}) The initial separation $r_0$ is $10^{-4}$ and the data are averaged over 10000 different simulations. One can distinctly note the presence of both the R-O law and the Batchelor's scaling law. The cross-over time $t_0$ is somewhere in the third decade on the x-axis. The deviation of the blue curve (marked $\left\langle|\textbf{r}(t) - \textbf{r}_{0}|^2\right\rangle/t^3$) from R-O scaling at higher times is due to the fact that the pair-separations are no longer in the inertial regime. All the units are in GOY-shell-model units.}\label{fig2}
\end{figure}

As a main result from our simulations, we find that both Batchelor's and R-O scalings are clearly observed, as shown in Fig. (\ref{fig2}). Please note that we have plotted $\left\langle|\textbf{r}(t) - \textbf{r}_{0}|^2\right\rangle$ and not $\left\langle|\textbf{r}(t)^2 - \textbf{r}_{0}^2|\right\rangle$ on the y-axis. This is to make contact with the experimental results \cite{Ouellette} which demonstrate that the correlation between the initial separation and the relative velocity of the pair may not be neglected when the flow is not perfectly homogeneous.
%\cite{Salazar}
However, the GOY model--- by construction--- is a shell model of a perfectly homogeneous and isotropic turbulence. So, within the paradigm of shell models, probably it doesn't make much of a difference if one chooses to work with $\left\langle|\textbf{r}(t)^2 - \textbf{r}_{0}^2|\right\rangle$.\\

%%%%THE FIGURES%%%%%%%%%%%%%%
\begin{figure}[hbt]
\includegraphics[width=0.50\textwidth]{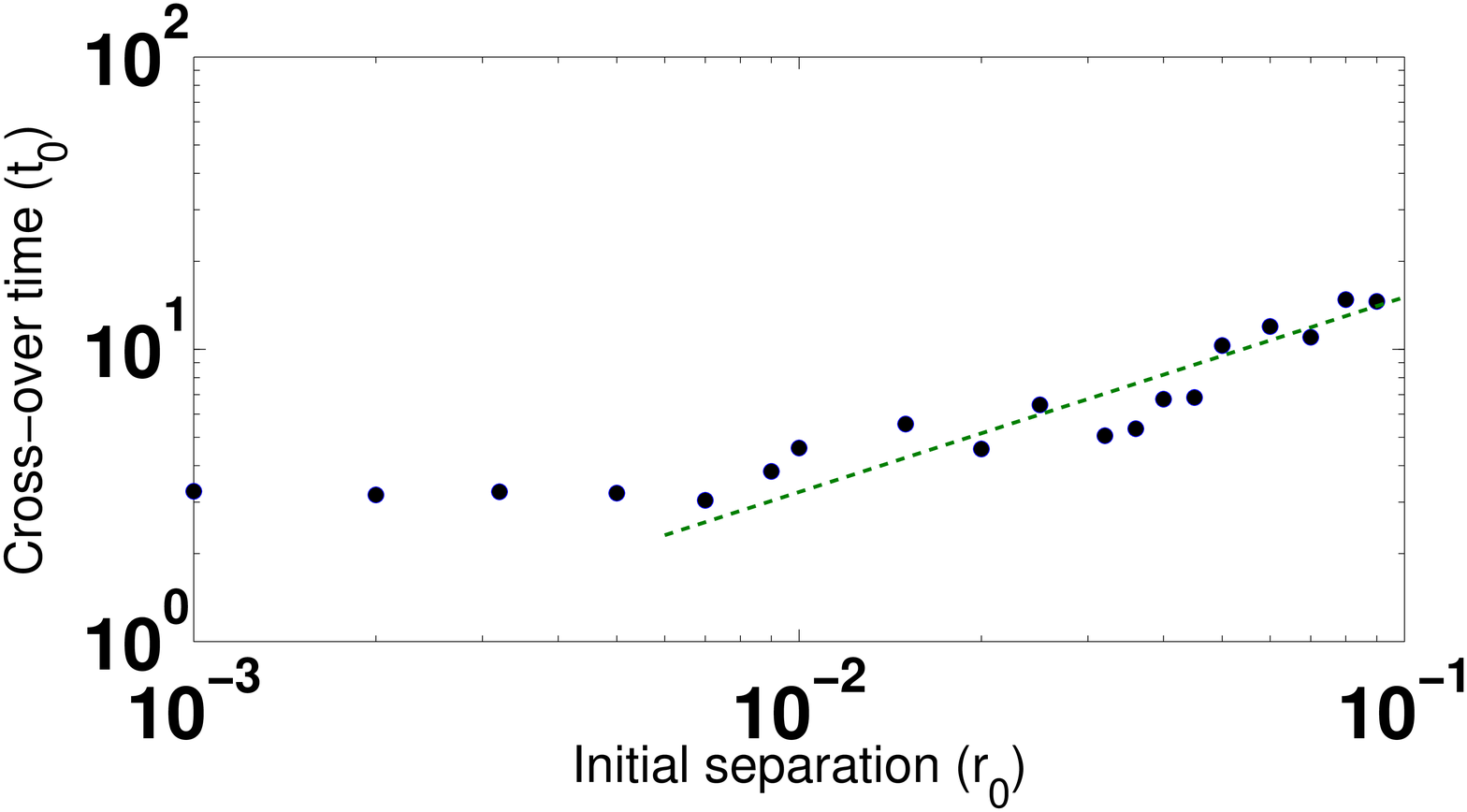}
\caption{({\it Colour online}) The cross-over times $t_0$ plotted the initial pair-separations $r_0$. The dashed line has the theoretical slope predicted by Batchelor. All the units are in GOY-shell-model units.}\label{fig3}
\end{figure}
%%%%THE FIGURES%%%%%%%%%%%%%%

With no knowledge of the exact functional dependence of $\left\langle|\textbf{r}(t) - \textbf{r}_{0}|^2\right\rangle$ on various parameters of the flow, it is quite a tough task to exactly locate the cross over time, $t_0$, between the Batchelor regime and the R-O regime. However, we can use the following rather crude but effective method to see whether equation (\ref{eq2}) is validated by our simulations.
To investigate where the crossover happens, we estimated the slope along the blue curve, $\left\langle|\textbf{r}(t) - \textbf{r}_{0}|^2\right\rangle/t^3$ in Fig. (\ref{fig2}). Obviously, when the slope is close to zero, we are in R-O scaling regime. As the crossover occurs before the slope is zero, we decide on the following algorithm: determine the time corresponding to the point where the slope is equal to $-0.5$\footnote{It may be mentioned that while experimenting with this approximate method of locating crossover time, we noted that results look more or less same for other values for ``cut-off" slope around $-0.5$; thus we report herein results only for the cut-off slope$=-0.5$.} and simply denote that $t_0$; repeat the procedure for the curves with various different $r_0$. We have used the data generated this way to obtain Fig. (\ref{fig3}) where we plot the cross-over times $t_0$ versus the initial pair-separations $r_0$. One can note the reasonable agreement of the data with the theoretical Batchelor prediction (the dashed line) when the initial separation
is not too small. What has been shown is that our data follow the expected trend; of course, data points are scattered around the theoretical slope.
The deviation at smaller initial separations --- between $r_0=10^{-3} \textrm{ to } r_0=5 \times 10^{-3}$ --- probably has the following explanation.
In the case of our numerics, the Kolmogorov scale, $\eta$, is of the order of $10^{-5}$.
The Reynolds number, $\textrm{Re}\sim 1/\nu$, is of the order of $10^{9}$.
One may thus note that the Taylor microscale, $\lambda\sim \textrm{Re}^{1/4}\eta$ is more than $100$ times greater than the Kolmogorov dissipation scale.
This leads to the fact that in our model $\lambda\sim 10^{-3}$, implying that the data points in the flat region of Fig. (\ref{fig3}) do not quite correspond to the initial separations which are {\textit {deep}} inside the inertial regime; and hence, it is not surprising that they do not follow the Batchelor prediction.
Nevertheless, we feel it is quite remarkable that the GOY shell-model of turbulence can reproduce this law very well for larger separations.
Similar result is obtained if one repeats the aforementioned algorithm for the red curve,$\left\langle|\textbf{r}(t) - \textbf{r}_{0}|^2\right\rangle/t^2$,
 in Fig. (\ref{fig2}) focusing on the Batchelor's scaling regime.
\\
Another way to find the relation between the $r_0$ and $t_0$ comes from the fact that R-O scaling is $\left\langle r^2\right\rangle = g \epsilon t^3$ while Batchelor considered $\left\langle|\textbf{r}(t)^2 - \textbf{r}_{0}^2|\right\rangle$. So, if we plot $\left\langle r^2\right\rangle$ and $\left\langle|\textbf{r}(t)^2 - \textbf{r}_{0}^2|\right\rangle$ versus time, they will become indistinguishable when R-O scaling sets in and this will happen around $t_0$ for the corresponding $r_0$. Thus, one can conveniently devise an algorithm to find $t_0$ in this case. We may mention that also from applying this indirect method, Eq. (\ref{eq2}) stands validated.
\\
In the passing, it may be mentioned that within the framework adopted in this paper, one can also study exit-time statistics\cite{boffetta1,boffetta2} of the pair dispersion phenomenon to good effect. The corresponding results will be reported elsewhere.

\section{Conclusion}
We have studied pair dispersion in a turbulent flow, applying the GOY shell model (Eq. (\ref{un})) and Fourier transforming the complex shell-velocities back into real space (Eq. (\ref{field})). This procedure results in strongly turbulent velocity field where the dispersion of pair of particles can be easily studied by advecting the passive particles in the velocity field. In particular, we have investigated how the dispersion depends on the initial separations.\\
We may mention that this is probably for the first time that a shell-model of turbulence has shown the simultaneous existence of both the R-O and the Batchelor regimes in turbulent pair dispersion.
Whereas devising a better method of finding $t_0$ would be worth investigating, the temporal extent over which these laws are valid can easily be increased if one uses a larger number of shells.\\
In closing, we hope that our results may spark interest in implementing GOY model to study pair dispersion in the enstrophy cascade dominated regime in two-dimensional flows. Also, it might be fruitful to use GOY model modified appropriately \cite{Sagar4} for investigating turbulent relative dispersion in more realistic case of rotating turbulence which is of profound geophysical and astrophysical interest.
\begin{acknowledgements}
We wish to thank Prof. R. Benzi for pointing out the importance of Taylor microscale in explaining our numerical results.
\end{acknowledgements}
%%%%%%%%%%%%%%%%%%%%%%%%%%%%%%%%%%%%%%%%%%%%%%%%%%%%%%%%%%%%%%%%%%%%%%%%%%%%%%%%%%%%%%%%%%%%%%%%%%%%%%%%%%%%%%%%%%%%%%%%%%%%

%%%%%%%%%%%%%%%%%%%%%%%%%%%%%%%%%%%%%%%%%%%%%%%%%%%%%%%%%%%%%%%%%%%%%%%%%%%%%%%%%%%%%%%%%%%%%%%%%%%%%%%%%%%%%%%%%%%%%%%%%%%%
\end{document}